\documentclass[a4paper,11pt]{article}
\pdfoutput=1 

\usepackage{jheppub} 

\usepackage[T1]{fontenc} 
\usepackage{grffile}
\usepackage{braket}
\usepackage{subcaption}
\usepackage[compat=1.0.0]{tikz-feynman}
\title{Correction to black hole radiation due to pair annihilation}

\author{Moslem Shafiee}

\affiliation{Department of Physics, Sharif University of Technology,\\
P.O. Box 11155-9161, Tehran, Iran}

\emailAdd{shafiee\_moslem@physics.sharif.ir}

\abstract{We consider the emission of charged scalar particles from a Schwarzschild black hole. It is shown that these particles can interact with each other through pair annihilation, and as a result, produce photons. These photons make a correction to the spectrum of photons that are directly emitted from the black hole. By solving field equations using the WKB approximation, the pair annihilation rate is taken into account for the most probable case, i.e., the first order, and therefore the correction will be of the order of $e^2$. Considering this scenario, we show that most interactions take place near the event horizon of the black hole, yet the number of interactions is not large enough to make a remarkable modification in the radiation spectrum of particles. These results refute the claims that interactions cause immense changes in the Hawking radiation spectrum of a black hole.}

\begin{document} 
\maketitle
\flushbottom

\section{Introduction}
Quantum theory predicts that black holes can emit thermal radiation due to the unique properties of their event horizon. The temperature of this radiation is inversely proportional to the mass of the black hole so that the lower the mass of the black hole, the more intense the radiation \cite{a, bh explosion}. With time progression and consequently increasing the temperature, the process of black hole evaporation occurs more quickly. Therefore, quantum effects in black holes which are small enough will be significant. In the case of an uncharged and non-rotating black hole, the type of emitted particles depends only on its mass. For example black holes with $M>>10^{17}g$ can only produce light and massless particles such as neutrinos, photons, and gravitons. But with condition $10^{14}g<<M<<10^{17}g$, the emission of particles like electrons and positrons is also allowed and a considerable part of the spectrum is allocated to them \cite{b}. Also for a black hole of $M<5\times10^{14}g$, composite particles such as hadrons contribute to Hawking radiation and even for smaller black holes, quarks and gluons can also be produced \cite{c, d}. In the other words, by increasing the black hole temperature the emission of heavier particles is more possible.

Heckler has claimed \cite{e, f} that when the black hole temperature exceeds 45 GeV, then the emitted Hawking radiation can sufficiently interact with itself and forms a photosphere around the black hole. Heckler has shown that dominant interactions are Bremsstrahlung and electron-photon production, and the photosphere is formed at the distance of $\alpha^{-4}r_s$ from the black hole ($\alpha$ is the fine structure constant and $r_s$ is the Schwarzschild radius), Consequently, the Hawking radiation spectrum changes considerably. Also, if the temperature of the black hole is greater than $\Lambda_{QCD}$ (the QCD scale parameter), then QCD interactions can occur, which causes the chromosphere formation around the black hole. In \cite{g} using the numerical solution of the Boltzmann equation by relaxation time approximation, the authors have investigated  QED and QCD interactions for the formation of  photosphere and chromosphere. They have demonstrated that the photosphere is formed at a lower temperature than Heckler's  predicted value, but the large modification in the Hawking spectrum still occurs. In \cite{h, i, j, k, l}, the Heckler model has been applied to predict the radiant spectrum of the primordial black holes for different types of particles, and the observational results are also considered, which all confirm Heckler's scenario.

But Page, Carr, and MacGibbon have shown that the photosphere creation would be impossible by taking into account two points that Heckler did not consider in his calculations \cite{m, n}. First, particles for doing interactions must be related causally to each other, which is true only for particles that are sufficiently close to the black hole, in which the number of these particles is very small. Second, a particle (electron) needs a distance of about $\frac{E}{m_e}$ to complete a Bremsstrahlung interaction, in which case the likelihood of such an interaction would be very limited. In fact, in the interaction between two electrons, the electromagnetic field of one of them must reach the other, which requires time, and few electrons satisfy this condition. It has been studied that Bremsstrahlung is possible through eight cases, of which inner Bremsstrahlung is more effective, but in general, all of these interactions will not have a significant effect on the Hawking spectrum of a black hole, which makes the proposed model in \cite{e,f} invalid. It has also been discussed with a similar argument that QCD interactions between particles emitted from black holes with a temperature of more than $\Lambda_{QCD}$ can be ignored with a very good approximation, and thus the chromosphere formation is canceled. They have argued that interactions, whether of QED or QCD, between particles emitted from a black hole, do not play a crucial role in modifying the Hawking radiation spectrum of the black hole.

In section \ref{bgf}, we will see that some of the produced particles fall into the black hole again,  because the space-time curvature near the horizon acts as a barrier against the particles, and only a fraction of them succeed to escape to infinity. In order to determine the distribution of particles in infinity, we first solve the field equations in Schwarzschild's background for bosonic particles by the WKB approximation. It is obtained that the distribution depends on the characteristics of the particle and the black hole. One of the most important features is the angular momentum of the particle that with its growth, the probability of the particle's escape from the black hole decreases. Then, we examine the distribution of particles with different spins, and by comparing their distribution with each other, we conclude that by increasing the spin, the particle contribution in the Hawking spectrum becomes less.

In the first part of section \ref{chr}, we look at the interactions in the presence of the black hole. The photon can be created in different ways from the interaction between emitted particles from a black hole, which the case under consideration here is the pair annihilation. For this purpose, we examine black holes with a mass of about $10^{15}g$, because a significant part of their spectrum belongs to the electrons and the positrons, and thus the pair annihilation  rate can be considered for them. Also, such black holes have a lifetime about the present age of the universe, and if they are formed in the early universe, then the final stages of their evaporation are running out now. For simplicity, we ignore the spin of electron and positron, and we actually use the scalar QED for the charged scalar particles (Considering their spin, the probability of their interaction also decreases as we will see in the next section, so the correction, in this case, will be smaller than what has been calculated through the scalar QED). Then, applying the usual methods of quantum field theory in curved space-time, we find a relation for the electron-positron annihilation and turning them into photons. Of course, we focus our attention on the state of the highest probability, i.e., the first order (tree level). It is demonstrated that this relation is a function of the particle's characteristics, such as angular momentum. By increasing the angular momentum of each of the particles participating in the interaction, the possibility of photon production is also reduced.

The second part of section \ref{chr} is devoted to the study of the pair annihilation impact on the photon spectrum. With the help of the obtained relation for the pair annihilation rate, we find a correction to the photon spectrum, and we will see that this correction which is the order of $e^2$ contains two parts. In the end, it is observed that due to the small number of interactions, the radiation spectrum of the black hole does not modify considerably, however, the interactions cause the spectrum to lose its completely thermal state.

In this paper, we use the natural units in which $G=\hbar=c=1$.
\section{Bosonic greybody factors}
\label{bgf}
An uncharged and non-rotating black hole with mass $M$ is described by the metric
\begin{equation}
ds^2=-\big(1-\frac{2M}{r}\big)dt^2+\big(1-\frac{2M}{r}\big)^{-1}dr^2+r^2(d\theta^2+\sin^2\theta d\varphi^2),\label{2.1}
\end{equation}
and the distributation for produced particles with integer spin  by this black hole is \cite{a}
\begin{equation}
N(\omega)=\frac{1}{e^{8\pi M\omega}-1},\label{2.2}
\end{equation}
where $\omega$ is the particle's energy. This relation is the same as the Planck distribution for blackbody radiation, with this assumption that the black hole has a temperature equal to $\frac{1}{8\pi M}$. Equation \eqref{2.2} describes the number density of emitted particles near the black hole event horizon. The number density in infinity varies with the above relation since some of the produced particles are again trapped by the black hole. In fact, the black hole radiation is not completely thermal and deviates from it. The distribution of particles far away from the black hole is given by the greybody factor, which depends on the characteristics of the particle and the black hole. The following is a way to find this factor.

\subsection{Quantum fields in Schwarzschild space-time}
Greybody factor is defined as the ratio of the energy flux in infinity to the energy flux on the event horizon. In order to calculate this factor, we consider the equation of a scalar field in \eqref{2.1} background
\begin{equation}
(\Box^{(4)}+m^2)\phi=0\label{2.3},
\end{equation}
where $\Box^{(4)}=g^{\mu\nu}\bigtriangledown_{\mu}\bigtriangledown_{\nu}$ is the four dimensional d'Alembertian. Solutions to this equation can be obtained through the method of separation of variables. By doing a bit of calculations, it can be shown that \cite{qft in cst}
\begin{equation}
\phi=\sum_{l=0}^{\infty} \sum_{m=-l}^{l}e^{-i\omega t}\frac{R_ l(r)}{r}\ Y_{lm} (\theta,\varphi),\label{2.4}
\end{equation}
which $R_ l(r)$ satisfies the below equation, using the tortoise coordinate $dr^*=\frac{dr}{(1-\frac{2M}{r})}$
\begin{equation}
\frac{d^2R_ l}{dr^{*^2}}+\Big{\{}\omega^2-\Big(1-\frac{2M}{r}\Big)\Big(\frac{2M}{r^3}+\frac{l(l+1)}{r^2}+m^2\Big)\Big{\}}R_ l=0.\label{2.5}
\end{equation}
\begin{figure}[t!]
\centering
\includegraphics[width=8cm]{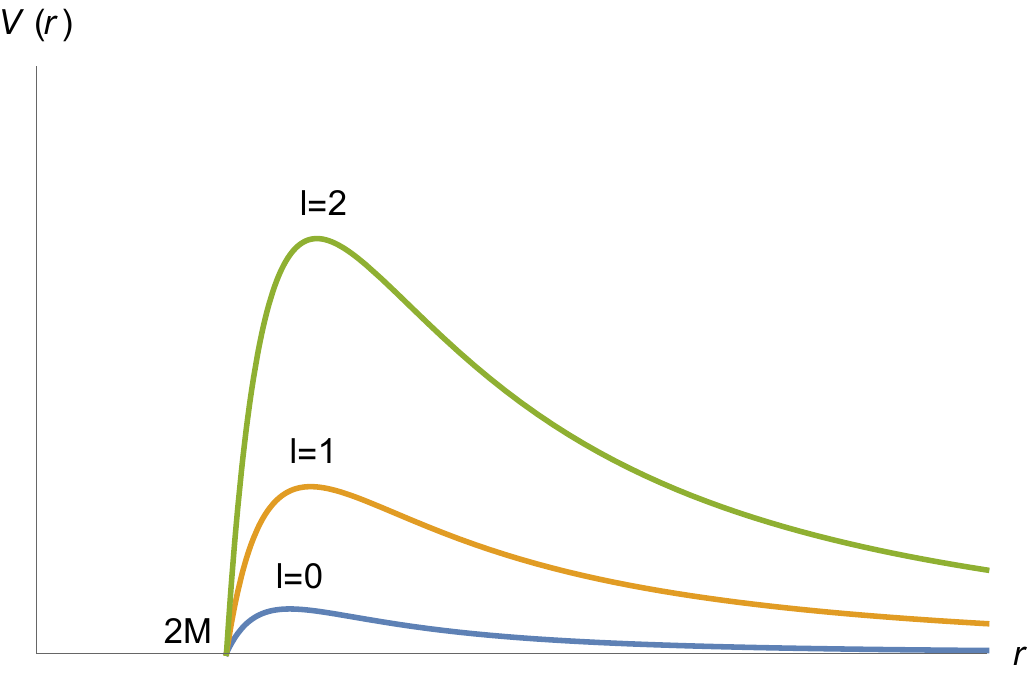}
\caption{Black hole potential for different values of angular momentum $l$.}
\label{bhp}
 \end{figure}
The following quantity is introduced as the potential of the black hole
\begin{equation}
V(r)=\Big(1-\frac{2M}{r}\Big)\Big(\frac{2M}{r^3}+\frac{l(l+1)}{r^2}+m^2\Big).\label{2.6}
\end{equation}
Equation \eqref{2.5} states that the produced particles on the horizon with energy $\omega$, angular momentum $l$, and with condition $V(r_{max})>\omega^ 2$ to escape to infinity must tunnel through the black hole potential barrier (figure \ref{bhp}). Therefore, some of the particles produced by the black hole fall into it again, and only a fraction of these particles survive from the gravitational attraction of the black hole through tunneling.

\begin{figure}[t!]
\centering
\includegraphics[width=8cm]{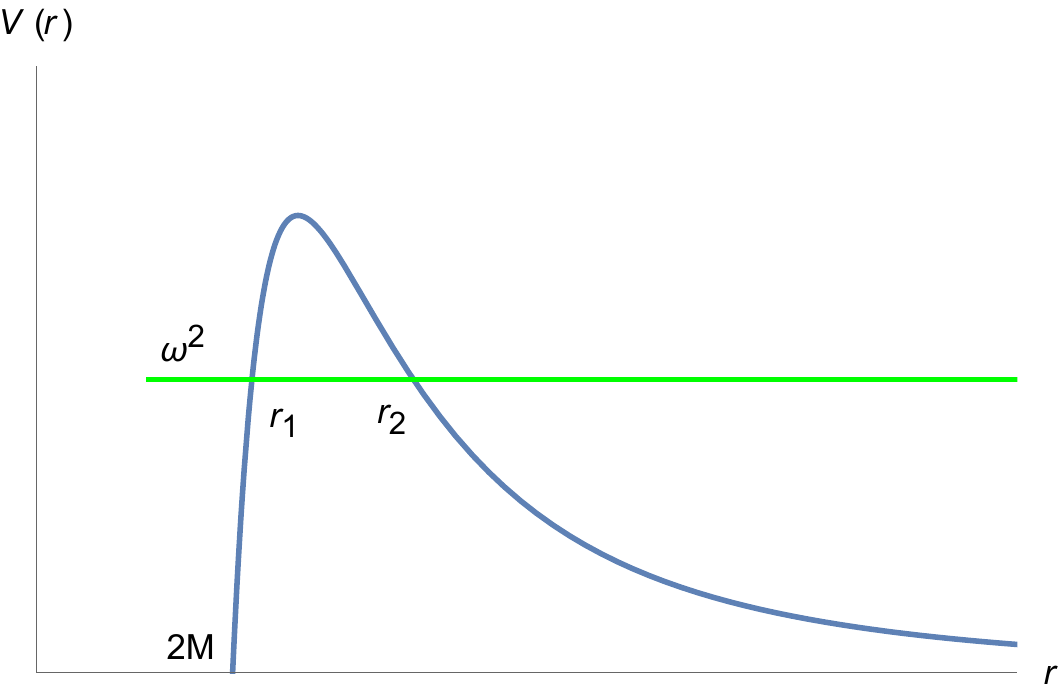}
\caption{Divided regions in \eqref{2.12}. A particle located in $2M<r<r_1$ must tunnel through the classically forbidden region $r_1<r<r_2$ to reach the area $r_2<r$.}
\label{rreq}
 \end{figure}

It has been demonstrated that equations of any massless field of spin $s$ in Schwarzschild geometry are reducible to the equation of a scalar field \cite{o, p} . In essence, the components of these fields can be obtained from solving the massless Klein-Gordon equation by replacing the potential associated with them as 
\begin{equation}
V_{sl}(r)=\Big{(}1-\frac{2M}{r}\Big{)}\Big{(}\frac{2M}{r^3}(1-s^2)+\frac{l(l+1)}{r^2}\Big{)},\label{2.7}\hspace{1cm} l\geq s.
\end{equation}
So it is just enough to solve \eqref{2.5} for $V_{sl}(r)$. To find the solutions to this equation, we use the WKB approximation method. Assuming the solutions of \eqref{2.5} are in the form of $R_{sl}=p(r)e^{iq(r)}$, and substituting it into the relation \eqref{2.5}, equaling the real and the imaginary parts to zero, separately, we get
\begin{align}
p^{''}+&pB_{sl}^{2}(r)-pq^{'2}=0,\label{2.8}
\\
&2p^{'}q^{'}+pq^{''}=0,\label{2.9}
\end{align}
where $B_{sl}^2(r)=\omega^2-V_{sl}(r)$. In \eqref{2.8}, if $p^{''}$  would not be too large, it can be ignored relative to the other two terms  \footnote{This\ condition\ comes\ from\ the\ WKB\ approximation\ method,\ which\ means\ that\ the\ changes\ in\ amplitude\\ $p$\ vary\ slowly,\ and\ this\ is\ only\ true\ when\ the\ energy\ of\ the\ particle\ is\ low.\ Thus,\ the\ WKB\ approximation\ is\ only\ valid\ for\ low\ energies\ regime.}. Thus with this assumption, we obtain from \eqref{2.8}
\begin{equation}
q(r)=\pm \int B_{sl}(r) d r^* \label{2.10},
\end{equation}
and by inserting the above value for $q$ in \eqref{2.9}, we find
\begin{equation}
p(r)=\frac{1}{\sqrt{|B_{sl}(r)|}}.\label{2.11}
\end{equation}
Therefore, the solutions of the radial equation \eqref{2.5} are
\begin{equation}
R_{sl}(r) = \left\{
\begin{array}{ll}
\frac{C_1}{\sqrt{B_{sl}(r)}}\ e^{i\int_{r}^{r_1} B_{sl}(r^*) d r^*}+\frac{C_2}{\sqrt{B_{sl}(r)}}\ e^{-i\int_{r}^{r_1} B_{sl}(r^*) d r^*} &  2M< r < r_1\\
\frac{C_3}{\sqrt{|B_{sl}(r)|}}\ e^{\int_{r_1}^{r} |B_{sl}(r^*)| d r^*}+\frac{C_4}{\sqrt{|B_{sl}(r)|}}\ e^{-\int_{r_1}^{r} |B_{sl}(r^*)| d r^*} &  r_1< r< r_2\\
\frac{C_5}{\sqrt{B_{sl}(r)}}\ e^{i\int_{r_2}^{r} B_{sl}(r^*) d r^*} &  r_2< r
\end{array},\label{2.12} 
\right.
\end{equation}
where $r_1$ and $r_2$ are the positive roots of $B_{sl}^2(r)=0$ (figure \ref{rreq}). Equation \eqref{2.12} states that a part of the incoming wave is reflected after reaching the potential barrier, and the other part of it passes through tunneling. By calculating the tunneling rate, the greybody factor can be found, and this will be the subject of the next part.

Now, according to the obtained solution for the Klein-Gordon equation, the complex scalar field can be written as
\begin{multline}
\phi=\sum_{l, m}\int_{0}^{\infty}\frac{d\omega}{2\pi}\ \Big{[}a_{lm}(\omega)\ e^{-i\omega t}\ \frac{R_{0 l}(r)}{r}\ Y_{lm}(\theta,\varphi)\ \\+b_{lm}^{\dagger}(\omega)\ e^{i\omega t}\ \frac{R_{0 l}^{*}(r)}{r}\ Y_{lm}^{*}(\theta,\varphi)\Big{]},\label{2.13}
\end{multline}
where $a^\dagger$, $a$, $b^\dagger$ and $b$ are creation and annihilation operators for particles and antiparticles respectively, which are defined as
\begin{equation}
a_{lm}(\omega)\ket0=b_{lm}(\omega)\ket0=0,\hspace{0.7cm} \forall \ \omega, l, m.\label{2.14}
\end{equation}
These operators also satisfy the following commutation relations
\begin{equation}
\big{[}a_{lm}(\omega),a_{l^{\prime}m^{\prime}}^\dagger(\omega^{\prime})\big{]}=\big{[}b_{lm}(\omega),b_{l^{\prime}m^{\prime}}^\dagger(\omega^{\prime})\big{]}=\delta_{ll^{\prime}} \delta_{mm^{\prime}}\delta(\omega-\omega^\prime).\label{2.15}
\end{equation}
In the same way, the photonic field is 
\begin{multline}
A^\mu=\sum_{l, m, \lambda}\int_{0}^{\infty}\frac{d\omega}{2\pi}\ \varepsilon_{\lambda}^{\mu}\ \Big{[}c_{lm\lambda}(\omega)\ e^{-i\omega t}\ \frac{R_{1l}(r)}{r}\ Y_{lm}(\theta,\varphi)\ \\+c_{lm\lambda}^{\dagger}(\omega)\ e^{i\omega t}\ \frac{R_{1l}^{*}(r)}{r}\ Y_{lm}^{*}(\theta,\varphi)\Big{]},\label{2.16}
\end{multline}
where $c^\dagger$ and $c$ are creation and annihilation oprators for photons, and there is a similar commutation relation with \eqref{2.15} between them. The $\lambda$ index is related to the photon polarization, which can take values of 1 or 2. If the photon moves in the radial direction, then the polarization vectors can be considered as follows
\begin{equation}
\varepsilon_{1}^{\mu}=\frac{1}{\sqrt{2}}
\begin{pmatrix}
0\\
0\\
1\\
i
\end{pmatrix}
,\ \varepsilon_{2}^{\mu}=\frac{1}{\sqrt{2}}
\begin{pmatrix}
0\\
0\\
1\\
-i
\end{pmatrix},\label{2.17}
\end{equation}
where $\varepsilon_{1}^{\mu}$ and $\varepsilon_{2}^{\mu}$ represent the right-handed and the left-handed polarizations, respectively. These vectors satisfy the orthonormality condition
\begin{equation}
(\varepsilon_{\lambda}^{\mu})^\dagger (\varepsilon_{\lambda^\prime}^{\mu})=\delta_{\lambda \lambda^\prime}.\label{2.18}
\end{equation}
It should be noted that in this section we only considered the outgoing modes, and in the following we have done all the calculations for these modes.
\subsection{Tunneling rate}
In order to determine the tunneling rate, i.e., the greybody factor for potential \eqref{2.7}, we must find a relation between the coefficients of solutions \eqref{2.12}. As one can see, these solutions are singular in  $r_1$ and $r_2$, because in the vicinity of these points the value of $p^{''}$ increases, and so it can no longer be neglected. Thus, to find a relation between $C_1$, $C_2$, $C_3$, $C_4$, and $C_5$, and consequently the transmission coefficient, we must solve the radial equation in the vicinity of these points  exactly. 
By accomplishing these items (see appendix \ref{rbc}), one can obtain the transmission coefficient or the greybody factor as
\begin{equation}
\Gamma_{sl}(\omega)=\frac{|C_5|^2}{|C_1|^2}=\frac{e^{-2\gamma_{sl}}}{(1-\frac{e^{-2\gamma_{sl}}}{4})^2},\label{2.34}
\end{equation}
\begin{figure}[t]
\centering
\begin{subfigure}[t]{0.48\textwidth}
 \includegraphics[width=6.77cm]{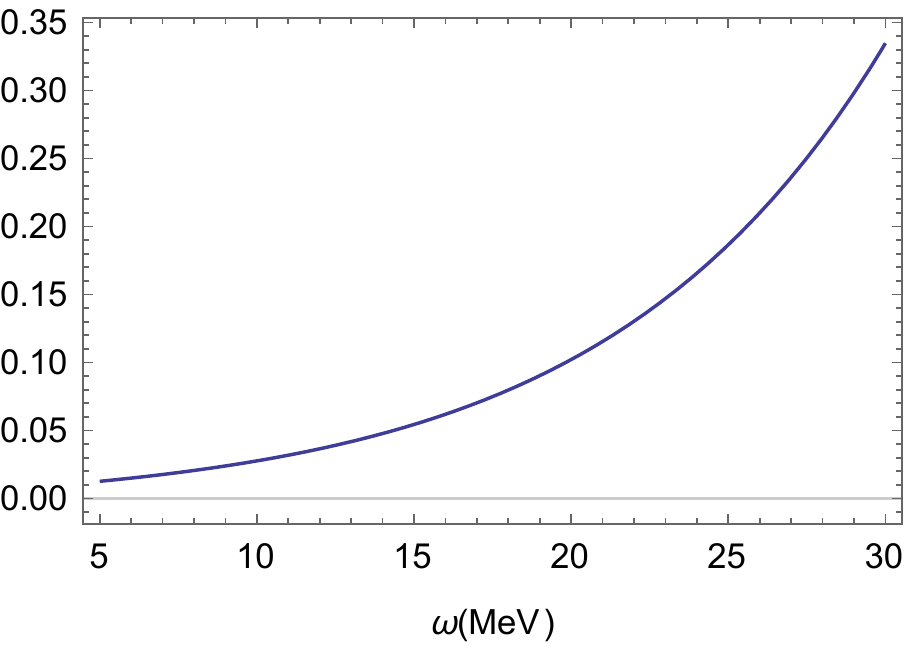}
\caption{s=l=0.}
\label{3a}
\end{subfigure}
~
\begin{subfigure}[t]{0.49\textwidth}
        \includegraphics[width=7cm]{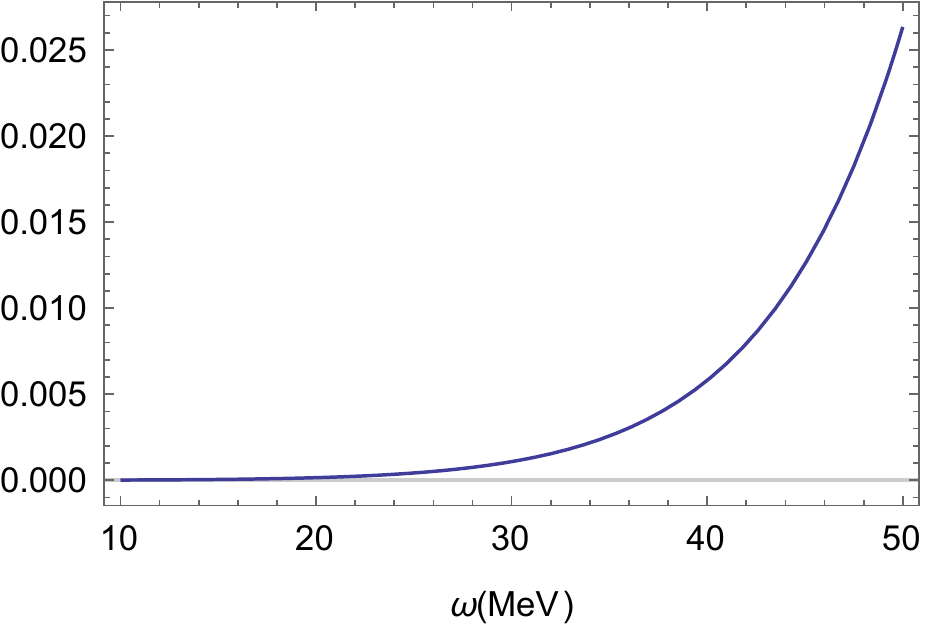}
        \caption{s=0, l=1.}
\label{3b}
\end{subfigure}
~
\begin{subfigure}[t]{0.48\textwidth}
 \includegraphics[width=6.77cm]{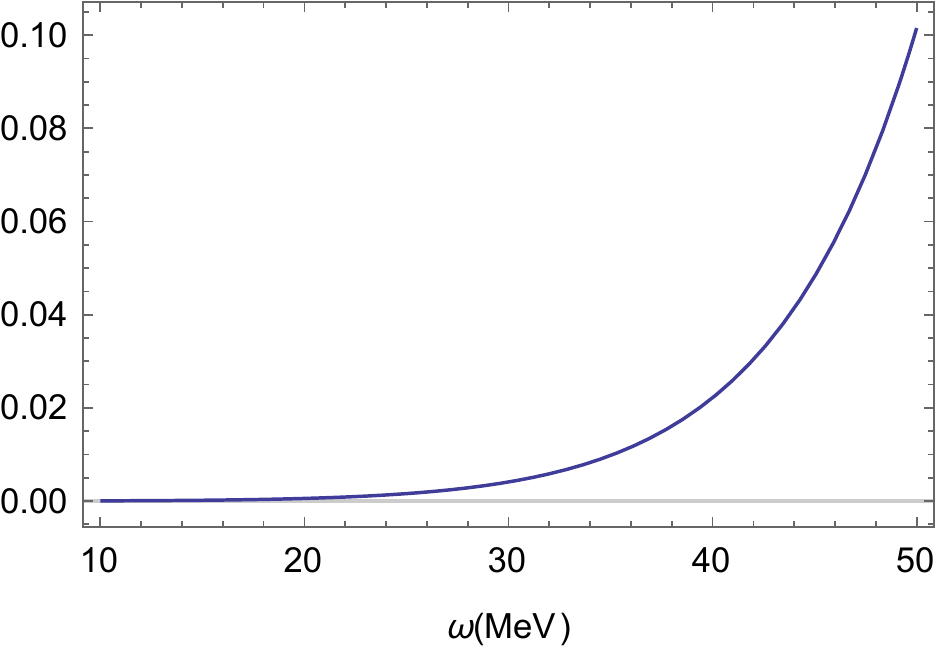}
\caption{s=l=1.}
\label{3c}
\end{subfigure}
~
\begin{subfigure}[t]{0.48\textwidth}
 \includegraphics[width=7.1cm]{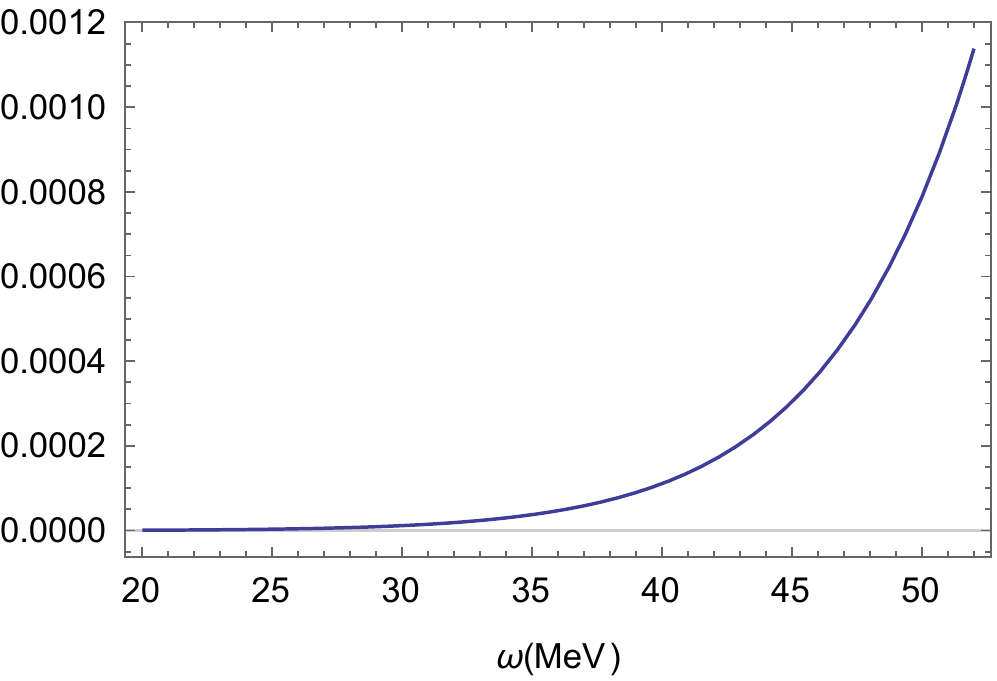}
\caption{s=l=2.}
\label{3d}
\end{subfigure}
 \caption{The greybody factors for the emitted massless bosons from a black hole of $10^{15}g$.}
\label{fig 3}
\end{figure}
which indicates that this coefficient depends on the energy, angular momentum, and spin of the particle, as well as on the black hole mass.  In this equation, for low energies (when the difference between $r_1$ and $r_2$ grows as it is seen in figure \ref{rreq}), the value of $\gamma_{sl}$ increases, which in this case the greybody factor with a very good approximation is equal to
\begin{equation}
\Gamma_{sl}(\omega)=e^{-2\gamma_{sl}},\hspace{1cm}  \gamma_{sl}>>0.\label{2.35}
\end{equation}
Of course, it is important to note that the relation \eqref{2.34} is not credible for the energies near $V_{sl}(r_{max})$. In fact, for these energies, the amount of \eqref{2.34} will be even greater than one, because the WKB approximation is valid only for low energies. So in order to avoid error in calculations, we set an upper limit for particle energy. For this purpose, we consider the energy $\omega_{sl}^{cut}=\Lambda_{sl}$ as the upper limit for which the relation \eqref{2.35}  is equal to $0.9$ times of \eqref{2.34}. This definition for $\Lambda_{sl}$ implies that $\gamma_{sl}(\Lambda_{sl})=\frac{1}{2}\ln5=0.80$. For example, the value of $\Lambda_{00}$ is about $30$ MeV. Therefore, we only look at a part of the spectrum, which the particle energy does not exceed $\Lambda_{sl}$. As we will see, this does not endanger the final conclusion. Further investigation for the validity of the WKB method for calculating the quasinormal modes of various black holes can be found in \cite{q}.

\begin{figure}[t!]
\centering
\begin{subfigure}[b]{0.48\textwidth}
 \includegraphics[width=7cm]{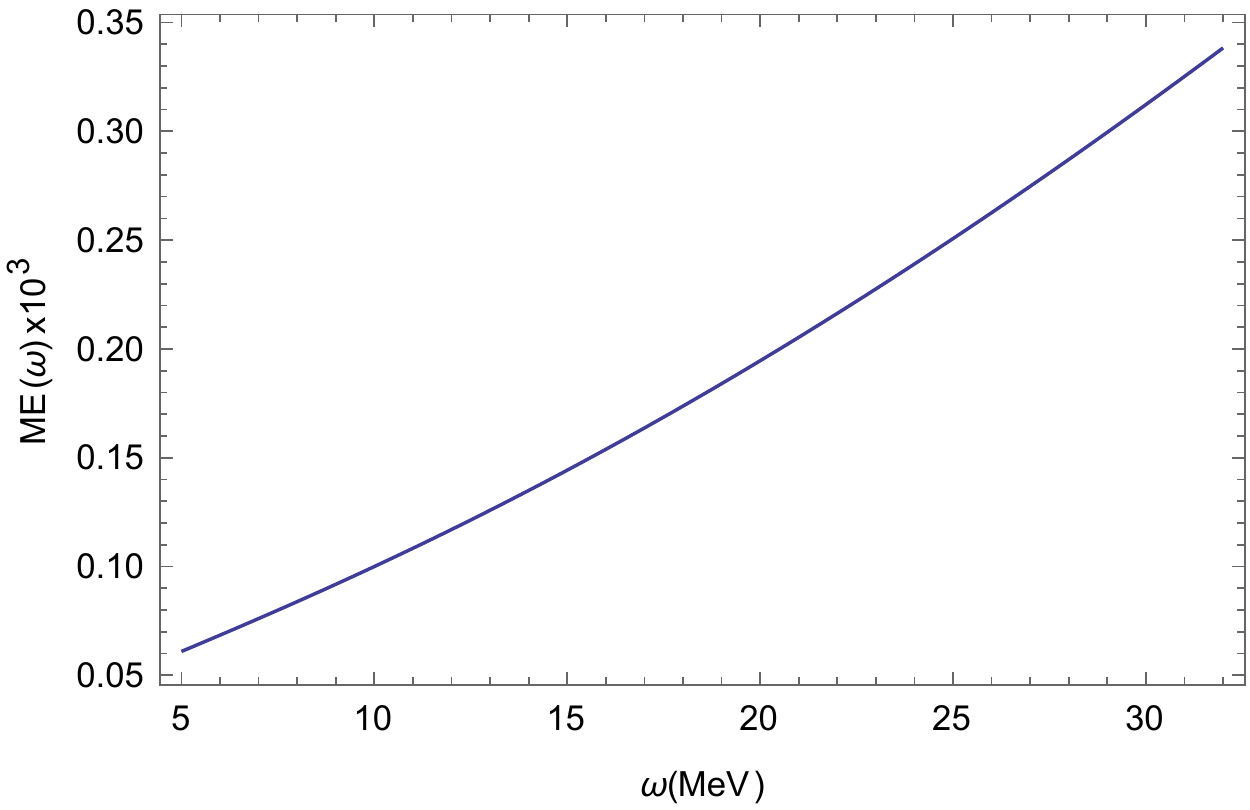}
\caption{s=l=0.}
\label{4a}
\end{subfigure}
~
\begin{subfigure}[b]{0.49\textwidth}
        \includegraphics[width=7.2cm]{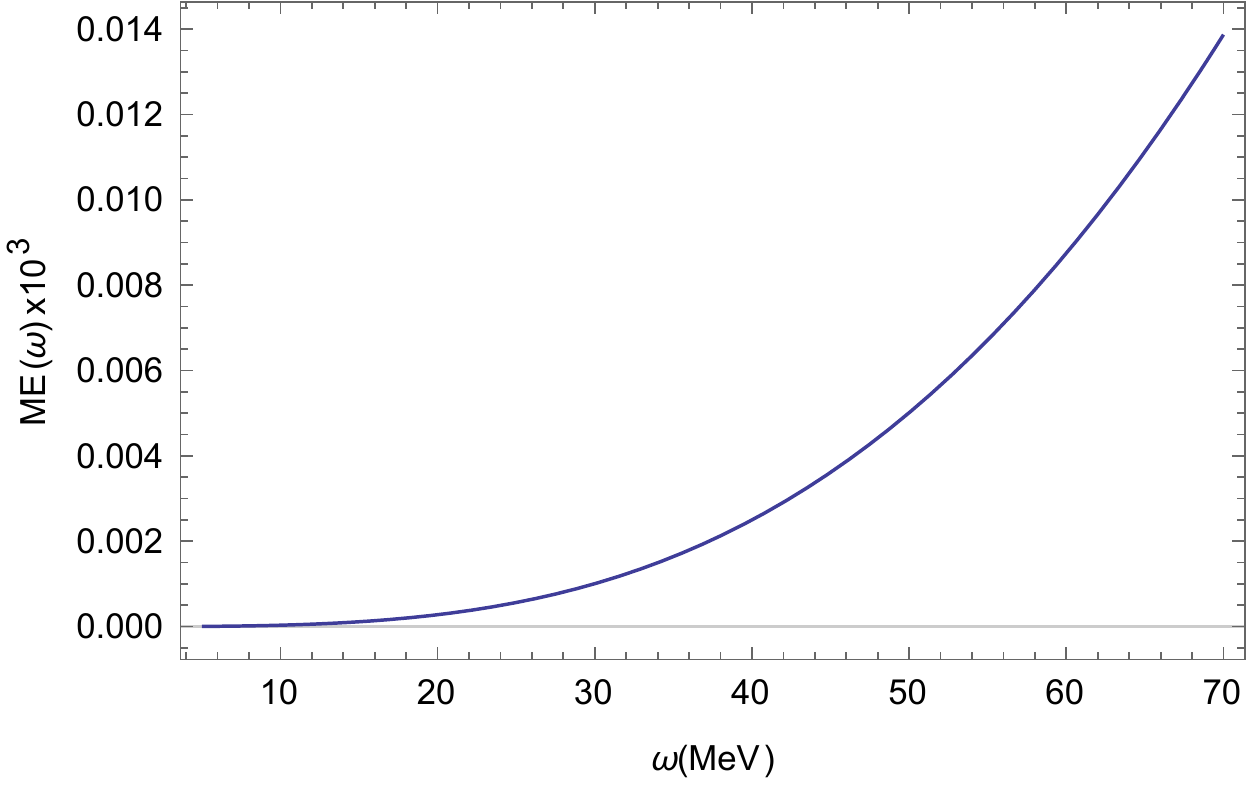}
        \caption{s=0, l=1.}
\label{4b}
\end{subfigure}
~
\begin{subfigure}[b]{0.48\textwidth}
 \includegraphics[width=7cm]{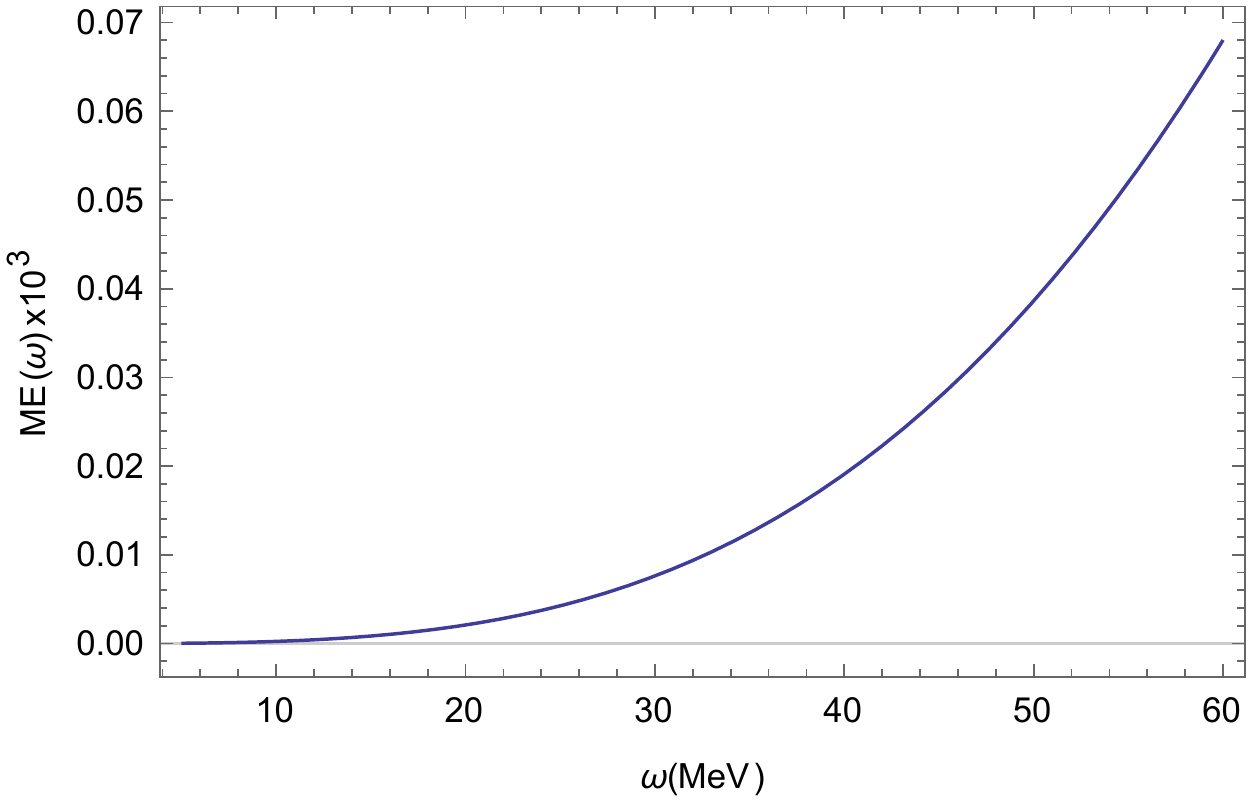}
\caption{s=l=1.}
\label{4c}
\end{subfigure}
~
\begin{subfigure}[b]{0.48\textwidth}
 \includegraphics[width=7.05cm]{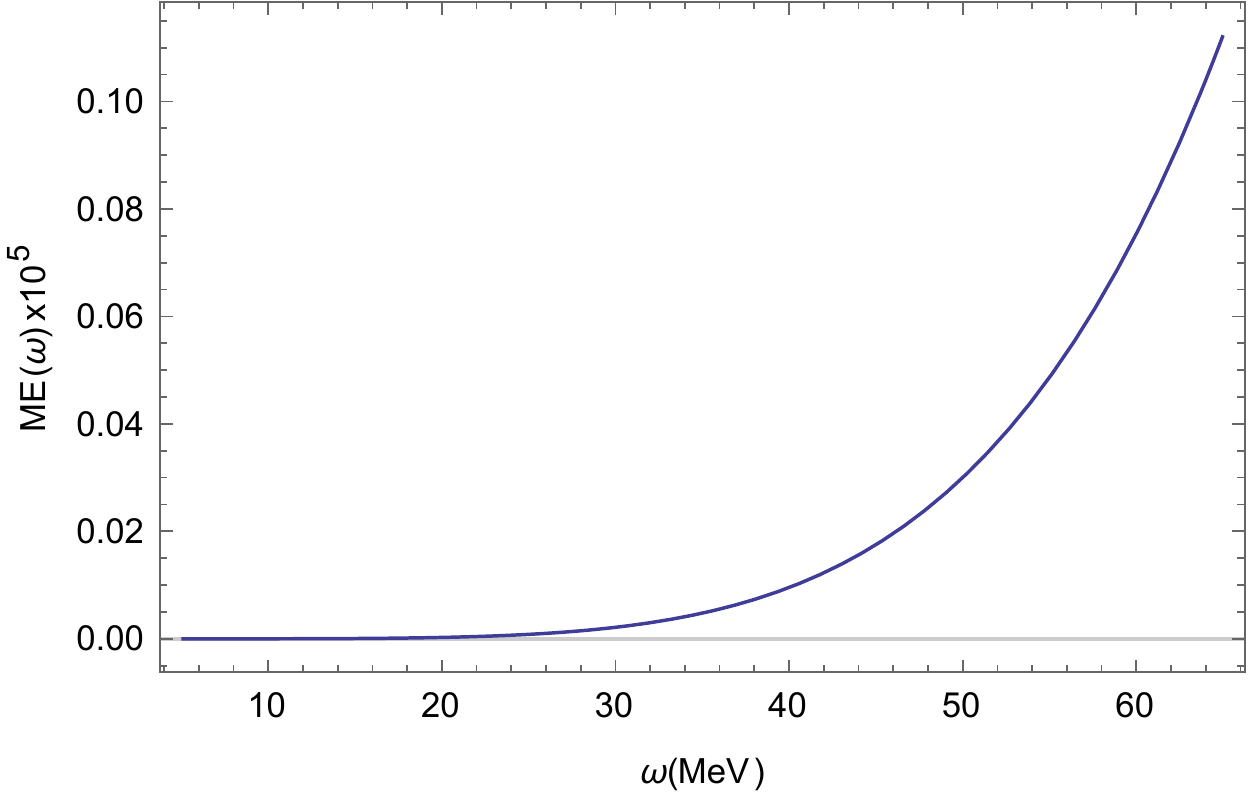}
\caption{s=l=2.}
\label{4d}
\end{subfigure}
 \caption{The energy flux for the emitted massless bosons from a black hole of $10^{15}g$ by taking into acount possible polarizations for photons and gravitons.}
\label{fig 4}
\end{figure}
In figure \ref{fig 3}, the value of the greybody factor for massless bosonic particles is shown. As it is seen in \ref{3a} and \ref{3b}, the value of the greybody factor for two particles with the same energy but different angular momentum is very distinct. As the angular momentum increases in accordance with figure \ref{bhp}, the height of the potential barrier grows, and hence the tunneling probability of the particle decreases. This causes that most of the spectrum of a spin $s$ particle is assigned to particles that have minimal angular momentum $s=l$. It can also be seen from the comparison of figures \ref{3b} and \ref{3c} that the value of the greybody factor for a photon is about ten times greater than a scalar particle with the same energy and angular momentum. The reason for this is the vanishing of the term which is proportional to $1-s^2$ in potential \eqref{2.7}, and hence the photon feels the less gravitational force, and it can tunnel easier. In fact, it can be said that a particle with equal spin and angular momentum $s$ has more contribution to the black hole radiation than a scalar particle with the same angular momentum. The greybody factor for different particles and black holes in various energy scales is calculated in \cite{b, r, s, t, u, v, w, x, y, z, aa, ab, ac, ad, ae, af, ag, ag1, ag2, ag3, ag4}.

Therefore, the number density of emitted particles in infinity is obtained
\begin{equation}
N_{sl}(\omega)=\frac{\Gamma_{sl}(\omega)}{e^{8\pi M\omega}-1},\label{2.36}
\end{equation}
and as well as the corresponding energy flux
\begin{equation}
E_{sl}(\omega)=\frac{1}{2\pi}\ \frac{\omega\Gamma_{sl}(\omega)}{e^{8\pi M\omega}-1}.\label{2.37}
\end{equation}
Figure \ref{fig 4} indicates the energy loss of a black hole due to the emission of different particles, for which the possible polarization states of photons and gravitons are also considered. It should be noted that the peak of the energy flux is not shown in figure \ref{fig 4}, since this value occurs for energies larger than $\Lambda_{sl}$, which we did not consider them due to the validity range of the WKB method. The dependence of the black hole potential on the particle characteristics causes the amount of energy flux to differ for each particle.
\section{Correction to Hawking radiation}
\label{chr}
The photon emission can be made in two ways, directly and indirectly from a black hole. Indirect radiation includes non-photonic emitted particles which can interact with each other and produce photons as a result. This can be a correction for the direct  emission of photons. Photon production is possible through various interactions, but the desired interaction here is pair annihilation. This correction according to figure \ref{rreq} consists of two parts. In the first part, particles and antiparticles collide together before reaching the potential barrier and create photons that are transmitted through tunneling to infinity. In the second part, the particle and the antiparticle first tunnel and then interact somewhere after the potential barrier, and ultimately the resulting photon reaches infinity.

In this section, we examine the interaction between  pairs of particles emitted from a black hole with a mass of about $10^{15}g$. The radiation spectrum of such a black hole is $45\%$ in electrons and positrons, $45\%$ in neutrinos, $9\%$ in photons, and $1\%$ in gravitons \cite{b}. In this case, the energy of most of the electrons and the positrons is ultrarelativistic that therefore their mass can be eliminated from the calculations. This black hole is large enough to consider the emission of electrons and positrons equally. The conditions for asymmetry in the emission of particles and antiparticles have been studied in \cite{ah}. Also, the evaporation rate in this black hole is slow enough to ignore the dependence of metric on the time. For simplicity, we ignore the spin of electron and positron in the calculations or, in other words, we use the scalar QED. The pair annihilation rate is computed in the first order, which causes the correction to be proportional to $e^2$.
\subsection{Interactions}
\begin{figure}
\centering
\begin{tikzpicture}
  \begin{feynman}
    \vertex (a) ;
    \vertex [above right=of a] (b);
    \vertex [above left=of b] (d) ;
    \vertex [right=of b] (c);
   
    \diagram* {
      (a) -- [fermion,edge label'=\(e^{-}\)] (b) -- [fermion,edge label'=\(e^{+}\)] (d),
      (b) -- [boson,orange] (c),
      
    };
  \end{feynman}
\end{tikzpicture}\hspace{3cm}
\begin{tikzpicture}
  \begin{feynman}
    \vertex (a) ;
    \vertex [above right=of a] (b);
    \vertex [above left=of b] (d) ;
    \vertex [above right=of b] (c);
 \vertex [below right=of b] (e);
   
    \diagram* {
      (a) -- [fermion,edge label'=\(e^{-}\)] (b) -- [fermion,edge label'=\(e^{+}\)] (d),
      (b) -- [boson,orange] (c),   (b) -- [boson,orange] (e), 
      
    };
  \end{feynman}
\end{tikzpicture}
\caption{Photon production in the first order through pair annihilation.}
\label{fig 5}
\end{figure}
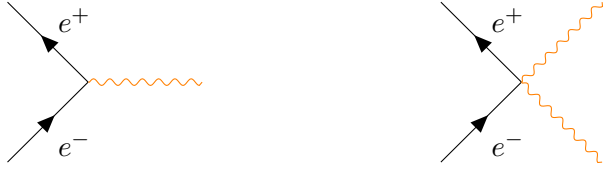

In scalar QED, interactions are discribed by the Lagrangian
\begin{equation}
\ell_{int}=ieA^\mu(\phi \partial_\mu \phi^*-\phi^*\partial_\mu \phi)+e^2A^2\phi\phi^*,\label{3.1}
\end{equation}
where the photonic and scalar fields are coupled together. If we consider the initial state as $\ket i$, then the transmission probability of  this state to the final state $\ket f$ is given by the following matrix
\begin{equation}
S=e^{i\int\ell_{int}\sqrt{-g}\ d^4x},\ \ S_{if}=\bra i S\ket f.
\end{equation}
By expanding the above relation, we have
\begin{equation}
S=I+i\int\ell_{int}\sqrt{-g}\ dx+\frac{(i)^2}{2}\big{(}\int\ell_{int}\sqrt{-g}\ d^4x\big{)}\big{(}\int\ell_{int}\sqrt{-g}\ d^4x^\prime\big)+\cdots
\end{equation}
Each term in this expansion represents different cases of state transition. In the first order, i.e., the second term in expansion, the photon production is possible via two diagrams for \eqref{3.1} (figure \ref{fig 5}). The states of  production of single and double-photon are respectively related to the first and second terms of \eqref{3.1}. It should be noted that in flat space-time, it is impossible to create a single photon from pair annihilation, but in curved space-time, this probability may not be zero. In fact, in this diagram, as we will see, the energy is conserved, but the conservation of momentum is no longer established because a part of the momentum in this diagram is absorbed by the black hole. In this case, the black hole acts as an external object in the interaction, something similar to that occurring in the Bremsstrahlung interactions.
Also, the single-photon production is of the order of $e^2$, while the state of double-photon production is of the order of $e^4$. So from now on, we only consider single-photon production.

Now assume that a particle and an antiparticle collide together with states $\ket{\omega_1,  l_1, m_1}$ and $\ket{\omega_2,  l_2, m_2}$, respectively, and create a photon with state $\ket{\omega, l, m}$. In this case, the fields of $\phi$ and $\phi^*$ must annihilate the particle and the antiparticle, and $A^{\mu}$ must produce the desired photon. Therefore, with respect to these explanations and \eqref{2.13} and \eqref{2.16}, one can show that the probability amplitude for this process is equal to
\begin{multline}
T_{l_1l_2l}\ (\omega_1,\omega_2,\omega)=-e\sqrt{\omega_1\omega_2\omega}\ \delta(\omega_1+\omega_2-\omega)\int_{r_{in}}^{r_{out}}\frac{dr}{r}\ \Big[R_{0 l_1}(r)R_{0 l_2}(r)R_{1l}^*(r)\Big]\\
\times\sum_{m_1,m_2}\sum_{m,\lambda}\int d\Omega\ \varepsilon_\lambda^\mu\Big(Y_{l_1m_1}(\partial_\mu Y_{l_2m_2})-Y_{l_2m_2}(\partial_\mu Y_{l_1m_1})\Big)Y_{lm}^*(\theta,\varphi).\label{3.4}
\end{multline}
The delta function denotes energy conservation. This equation implies that $m=m_1+m_2$, otherwise the integral value will be zero. As each of the $l_1$, $l_2$, or $l$ increases, the value of the angular part of the integral becomes smaller due to the appearance of higher powers of $\sin$ and $\cos$ functions in spherical harmonics. Also, with increasing $r$ and getting away from the black hole, the space-time becomes closer to the flat case, which makes the probability amplitude smaller. $r_{in}$ and $r_{out}$ represent the boundary of places where particles can interact with each other.

Thus if we suppose that $T^{(1)}$ and $T^{(2)}$ to be the values of $T_{l_1l_2l}$ for $(r_{in}=2M\ , r_{out}=r_1)$ and $(r_{in}=r_2\ , r_{out}=+\infty)$ respectively, then the corrected photon spectrum can be written as follows
\begin{multline}
N_{1l}(\omega)=\frac{2\Gamma_{1l}(\omega)}{e^{8\pi M\omega}-1}+\ \Gamma_{1l}(\omega)\sum_{l_1,l_2}\int_{m}^{\Lambda_{0l_1}}d\omega_1\int_{m}^{\Lambda_{0l_2}}d\omega_2\ \Big[\ \frac{1}{e^{8\pi M\omega_1}-1}\ \frac{1}{e^{8\pi M\omega_2}-1}\\ \times|T_{l_1l_2l}^{(1)}|^2\ \Big]
+\sum_{l_1,l_2}\int_{m}^{\Lambda_{0 l_1}}d\omega_1\int_{m}^{\Lambda_{0 l_2}}d\omega_2\ \Big[\ \frac{\Gamma_{0 l_1}(\omega_1)}{e^{8\pi M\omega_1}-1}\ \frac{\Gamma_{0 l_2}(\omega_2)}{e^{8\pi M\omega_2}-1}\ |T_{l_1l_2l}^{(2)}|^2\ \Big].\label{3.5}
\end{multline}
The first term represents the direct Hawking radiation of photons, which a factor of two appears due to the consideration of photon polarizations. The second and third terms express the indirect radiation of the photons (corrections), which are related to the pair annihilation before and after the potential barrier, respectively. $m$ is the rest mass of particle and antiparticle. The upper limits of the integrals are considered to be equal to $\Lambda_{0l_1}$ and $\Lambda_{0l_2}$, which indicate the validity range for the energy of particle and antiparticle, respectively. Note that in the second term the pair of particles before the potential barrier are annihilated and transformed into a photon, so it is the photon that sees the potential barrier in front of itself. Also, the value of $T^{(2)}$ is smaller than $T^{(1)}$, because the particles in this region are in a weaker gravitational field.  In the following, we will study the role of this correction in modifying the photon spectrum.

\subsection{Effect of interactions on the photon spectrum}
As we have found in the previous section, most of the spectrum of a spin $s$ particle belongs to particles with minimum angular momentum $s=l$. Accordingly, the correction is checked for the emission of photons with $l=1$. The highest probability of producing a photon of  $l=1$  is when one of the particles of the pair does not have angular momentum and the other has an angular momentum of  $l=1$ . Note that \eqref{3.4} is antisymmetric with respect to the to the displacement of $l_1$ and $l_2$, and therefore the probability value does not change with $l_1\longleftrightarrow l_2$. Also, with the increase of $l_1$ or $l_2$, as discussed in section \ref{bgf}, the number of particles decreases, and hence the number of interactions is reduced. Also, the number of interactions after the potential barrier decreases dramatically, since a large number of particles are reflected when they reach the potential barrier and fall into the black hole and, in addition, the fact that $T^{(2)}$ has a smaller value than $T^{(1)}$. This makes the second part of the correction much smaller than its first part. Thus, we only consider the larger part of correction, i.e., the second term in \eqref{3.5}.

\begin{figure}[t!]
\centering
\includegraphics[width=9cm]{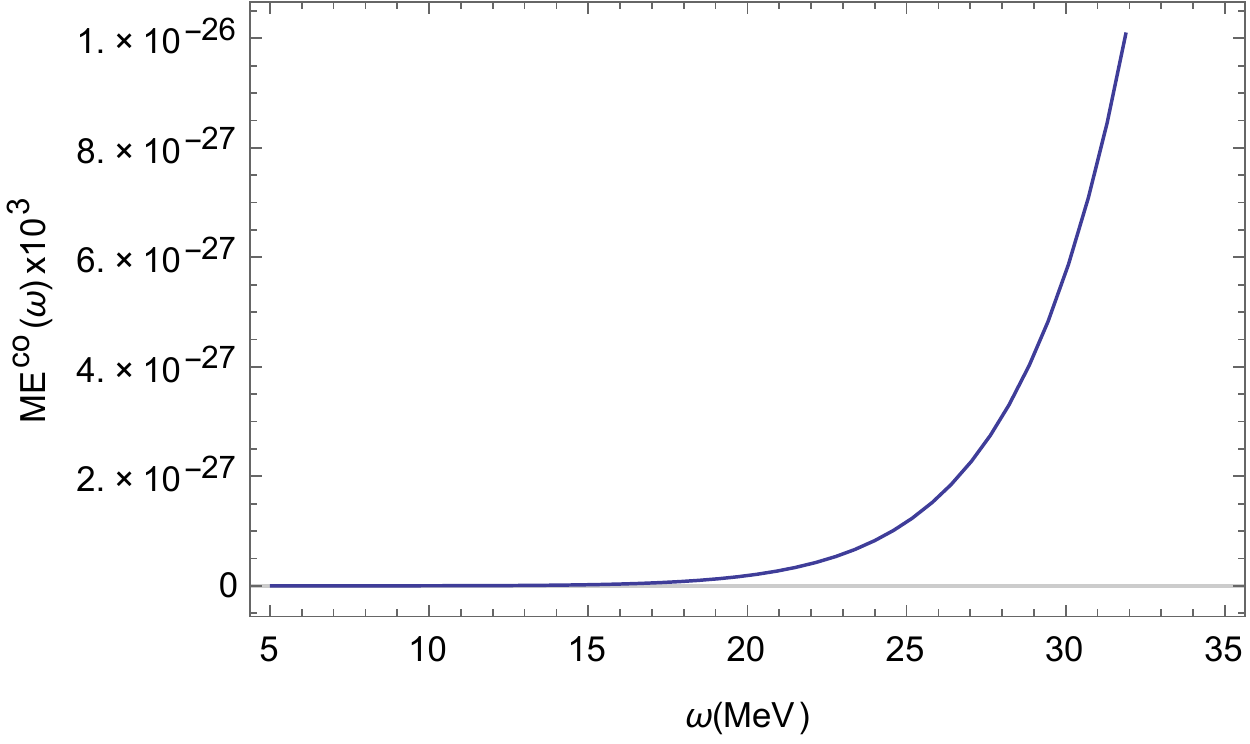}
\caption{ The corrected energy flux of photons with $l=1$ for a black hole of $10^{15}g$. }
\label{fig 6}
 \end{figure}

In oreder to calculate the desired correction, we need to apply an approximation. The potential of the black hole vanishes on the horizon. Also for energies of less than $\Lambda_{sl}$, according to figure \ref{rreq}, the value of $r_1$ is very close to $2M$ (as $\omega\rightarrow 0$ then $r_1\rightarrow 2M$). So with a very good approximation, we can ignore the potential in this region for particles with such energies. With this assumption, the solutions of \eqref{2.12} take the form
\begin{equation}
R_{sl}(r)=\frac{e^{-i\omega r^*}}{\sqrt{\omega}},\label{3.6}
\end{equation}
where, as mentioned before, we only consider the outgoing modes of the fields. Thus the value of $T_{101}^{(1)}$ equals to
\begin{multline}
T_{101}^{(1)}=-e\sqrt{\frac{1}{2\pi}}\ \sqrt{\omega_1\omega_2\omega}\ \delta(\omega_1+\omega_2-\omega)\int_{2M}^{r_1(\omega_{min})}\frac{dr}{r}\ \frac{e^{-i(\omega_1+\omega_2-\omega)r^*}}{\sqrt{\omega_1\omega_2\omega}}\\
=-e\sqrt{\frac{1}{2\pi}}\ \delta(\omega_1+\omega_2-\omega)\int_{2M}^{r_1(\omega_{min})}\frac{dr}{r}\ e^{-i(\omega_1+\omega_2-\omega)r^*},\label{3.7}
\end{multline}
which the factor $\sqrt{\frac{1}{2\pi}}$ is the value of the angular part of the integral. $\omega_{min}$ is the minimum between $\omega_1$ and $\omega_2$ because a particle which has lower energy will reach the potential barrier sooner. Eventually, after doing the related calculations, the probability value is obtained
\begin{equation}
|T_{101}^{(1)}|^2=\frac{e^2}{2\pi}\Big[\ln\frac{r_1(\omega_{min})}{2M}\Big]^2\ \delta(\omega_1+\omega_2-\omega)\label{3.8}
\end{equation}
As seen in the above equation, by increasing the black hole mass, the value of probability becomes less. Also by increasing $\omega_{min}$, the possibility of photon production enhances. In other words, photon production depends on two factors, the energy of the pair of particles and the mass of the black hole. Thus, the corrected energy flux for photons of $l=1$ can be written as
\begin{multline}
E_{11}^{co}(\omega)=\frac{e^2}{2\pi^2}\ \omega\Gamma_{11}(\omega)\int_{0}^{\Lambda_{00}}d\omega_1\int_{0}^{\Lambda_{01}}d\omega_2\ \bigg(\delta(\omega_1+\omega_2-\omega)\\
\times\frac{1}{e^{8\pi M\omega_1}-1}\ \frac{1}{e^{8\pi M\omega_2}-1}\Big[\ln\frac{r_1(\omega_{min})}{2M}\Big]^2\bigg).\label{3.9}
\end{multline}
By eliminating one of the integrals with the help of the delta function, we get
\begin{equation}
E _{11}^{co}(\omega)=\frac{e^2}{2\pi^2}\ \omega\Gamma_{11}(\omega)\int_{0}^{\Lambda_{00}}d\omega_1\ \bigg(
\frac{1}{e^{8\pi M\omega_1}-1}\ \frac{1}{e^{8\pi M(\omega-\omega_1)}-1}\Big[\ln\frac{r_1(\omega_{min})}{2M}\Big]^2\bigg).\label{3.10}
\end{equation}
In the extraction of the above relation, we consider that electrons and positrons to be massless since most of them are emitted ultrarelativisticly. Also, due to the symmetry of $|T_{l_1l_2l}|^2=|T_{l_2l_1l}|^2$ the equation is multiplied by a factor of two. The correction value for a black hole with a mass of about $10^{15}g$ is shown in figure \ref{fig 6}. Comparing this diagram with figure \ref{4c}, we find that the correction is much smaller than the direct emission of photons. In fact, this ratio is equal to
\begin{equation}
\frac{E_{11}^{co}(\omega)}{E_{11}(\omega)}\approx{10}^{-24}
\end{equation}
This indicates that the photons created by the pair annihilation are so rare. it is necessary to mention that the more we get away from the black hole, the number of interactions is reduced because the surface number density of particles is proportional to the inverse square of the distance. This means that in the distances away from the black hole, there are practically no interactions between the emitted particles. In fact, the number of interactions on the horizon is greater than anywhere else. Therefore, with a very good approximation, it can be said that the indirect radiation is due to the interaction of particles near the horizon. Finally, it can be concluded that pair annihilation does not play a significant role in modifying the photon spectrum. Of course, the interactions will remove the black hole radiation from the thermality and hence may carry the information of particles participating in the interactions. By subtracting the correction from the electrons and positrons emission, one can also find the amount of change in their spectrum. Similarly, It can be examined the other QED interactions between different particles emitted from a black hole. These interactions will be in the order of $O(e^2)$ which seems  to be very small for modifying the spectrum.  the first higher order for the single-photon production state  by pair annihilation occurs when the created photon decomposes into a particle and an antiparticle, and then the photon is formed again with their collision. The probability of this occurrence is proportional to $e^6$, which is much smaller than that discussed in this article. Thus, the loops have a much lower effect and, with a very good approximation, it can be said that the indirect Hawking radiation is due to the first order diagrams (figure \ref{fig 5}). Finally, the radiation spectrum of a black hole changes very slightly because of the interaction between the emitted particles, so it can be ignored. These results reject the proposed scenarios of \cite{e, f, g, h, i, j, k, l}, in which the interactions cause a considerable modification in the Hawking radiation spectrum of a black hole. However, the interactions take the radiation spectrum out of the thermal state and therefore carry information.
\section{Conclusion and remarks}
In this paper, we studied the effects of interactions on black hole radiation. We first obtained the emission rate from an uncharged and non-rotating  black hole. Various particles in this space-time feel different curvature and hence the emission rate depends on the type of particles. By solving the field equations through the WKB approximation, the tunneling rate from the black hole potential was obtained. It was determined that this value depends on the energy, angular momentum, and spin of the particle. As the angular momentum increases, the height of the potential barrier grows, and thus the tunneling of the particle becomes more difficult. This causes particles with a smaller angular momentum to have a larger contribution to the spectrum. Concerning the validity of the WKB method, we were only able to examine a part of the spectrum which is related to the low energies regime. In the following, it was shown that emitted particles can interact with each other and modify the Hawking radiation spectrum. By studying the emission of charged particles pair, it was observed that they can produce  photons through pair annihilation. These photons make corrections to the spectrum of photons created by the black hole. We calculated the amount of pair annihilation for the first order which has the highest chance of occurrence and we saw that it was proportional to $e^2$. Then we obtained the correction for the photonic spectrum with the help of the transmission equation of the pair to the photon. In order to produce photons from this process, the mass of the black hole and the energy of particles participating in the interaction play a crucial role. As the black hole mass increases, the possibility of photon formation will be reduced, while by increasing the energy of the particles, the probability value enhances. However, in general, the correction value is much smaller than the direct Hawking radiation of photons and therefore does not cause a significant modification in the spectrum.  Thus according to this point, it can be said that the interactions between the particles emitted from a black hole are negligible and with a good approximation, the spectrum of a particle is the same as the direct radiation of that particle. This result is also obtained from the examination of other interactions \cite{m, n}.

By taking into account the interactions in Hawking radiation, the radiation is no longer thermal. It has been proved that self-gravitation interactions exit the black hole radiation from a pure thermal state, which can be proposed as a solution to the information loss paradox \cite{ai}. These interactions also modify the Bekenstein-Hawking entropy, which confirms the unitarity evolution of black hole radiation \cite{aj}. But in this case, there is no correlation in Hawking radiation. However, adding corrections related to higher curvature in calculations of \cite{ai}, a correlation between the Hawking spectrum can be found \cite{ak}.

 It has also been argued that by considering loops of a $\lambda\phi^4$ interaction for a collapsing star, there is a correction that adds a term that is proportional to time to the energy flux. this causes the Hawking flux to be infinite by elapsing a long time \cite{al}. In other words, in this case, the perturbation theory breaks down, and to understand this process, one must sum all the loops together. This could be a sign that the picture in which black holes continue their thermal radiation until they fully evaporated is not correct. Thus for a better understanding of black hole radiation, the issue of the effect of self-interaction corrections on the Hawking radiation of a black hole requires further research.

\acknowledgments

The author appreciates Hessamaddin Arfaei for fruitful discussions and valuable comments.
\appendix
\section{The transmission coefficient}\label{rbc}
In this appendix, we solve the radial equation \eqref{2.5} near the points $r_1$ and $r_2$ exactly, to derive relation \eqref{2.34} for the transmission coefficient.

 When $r\approx r_1$, then the potential \eqref{2.7} takes the form
\begin{equation*}
V_{sl}(r)=V_{sl}(r_1)+(r^*-r_1)\frac{dV_{sl}}{dr^*}\Big|_{r=r_1}=\omega^2+(r^*-r_1)F_1,
\end{equation*}
where  $F_1=\frac{dV_{sl}}{dr^*}\big|_{r=r_1}$. By taking into account the potential as above and changing the variable $u=(r^*-r_1){F_1}^{\frac{1}{3}}$, the radial equation can be rewritten as
\begin {equation}
\frac{d^2R_{sl}}{du^2}-uR_{sl}=0.\label{2.19}
\end{equation}
The solutions to this equation are the first and second kind of Airy functions, Which are defined as follows respectively
\begin{equation*}
Ai(u)=\frac{1}{\pi}\int_{\circ}^{\infty} \cos(\frac{v^3}{3}+vu)  dv,
\end{equation*}
\begin{equation*}
Bi(u)=\frac{1}{\pi}\int_{\circ}^{\infty}\Big{[}e^{\frac{-v^3}{3}+vu}+\sin(\frac{v^3}{3}+vu)\Big{]}  dv.
\end{equation*}
So the general solution is obtained
\begin{equation}
R_{sl}(u)=aAi(u)+bBi(u).\label{2.20}
\end{equation}
To match this solution to \eqref{2.12}, we must use the asymptotic behavior of the Airy functions
\begin{subequations}
\begin{equation}
\left\{
\begin{array}{ll}
Ai(u)\approx \frac{1}{\sqrt{\pi} {|u|}^{\frac{1}{4}}}\ \sin\big{[}\frac{2}{3}{|u|}^\frac{-3}{2}+\frac{\pi}{4}\big{]} &  u<< 0\\
Ai(u)\approx \frac{1}{2\sqrt{\pi} u^{\frac{1}{4}}}\ e^{\frac{-2}{3}u^{\frac{3}{2}}} &  u>> 0
\end{array}
\right.\label{2.21a}
\end{equation}
\begin{equation}
\left\{
\begin{array}{ll}
Bi(u)\approx \frac{1}{\sqrt{\pi} {|u|}^{\frac{1}{4}}}\ \cos\big{[}\frac{2}{3}{|u|}^\frac{-3}{2}+\frac{\pi}{4}\big{]} &  u<< 0\\
Bi(u)\approx \frac{1}{2\sqrt{\pi} u^{\frac{1}{4}}}\ e^{\frac{2}{3}u^{\frac{3}{2}}} &  u>> 0
\end{array}
\right.\label{2.21b}
\end{equation}
\end{subequations}
Also $r\approx r_1$ implies that
\begin{align*}
&B_{sl}^2(r)=\omega^2-V_{sl}(r)=-(r^*-r_1)F_1\\
\\ \Rightarrow &\int_{r}^{r_1} B_{sl}(r^*) d r^*=\int_{u}^{0} |u^{'}| d u^{'}=\frac{2}{3}|u|^{\frac{3}{2}}.
\end{align*}
With the help of the above relation and the asymptotic behavior of the Airy functions, we can connect \eqref{2.20} to \eqref{2.12} for the case $u<0$
\begin{multline}
\frac{F_1^{\frac{1}{6}}}{\sqrt{\pi} \sqrt{B_{sl}(r)}}\ \Big{[}a \sin\big(\int_{r}^{r_1} B(r^*)  d r^*+\frac{\pi}{4}\big)+b \cos\big(\int_{r}^{r_1} B_{sl}(r^*) d r^*+\frac{\pi}{4}\big)\Big{]}\\=\frac{C_1}{\sqrt{B_{sl}(r)}}\ e^{i\int_{r}^{r_1} B_{sl}(r^*) d r^*}+\frac{C_2}{\sqrt{B_{sl}(r)}}\ e^{-i\int_{r}^{r_1} B_{sl}(r^*) d r^*}.\label{2.22}
\end{multline}
Therefore
\begin{align}
C_1=\frac{F_1^{\frac{1}{6}}}{2\sqrt{\pi}}\ e^{\frac{i\pi}{4}}(b+\frac{a}{i}),
\ \ C_2=\frac{F_1^{\frac{1}{6}}}{2\sqrt{\pi}}\ e^{\frac{-i\pi}{4}}(b-\frac{a}{i}).\label{2.23}
\end{align}
Similarly, it can be shown for $u>0$
\begin{equation}
C_3=\frac{bF_1^{\frac{1}{6}}}{\sqrt{\pi}},\ \  C_4=\frac{aF_1^{\frac{1}{6}}}{2\sqrt{\pi}}.\label{2.24}
\end{equation}
By comparing \eqref{2.23} and \eqref{2.24}, we finally get
\begin{subequations}
\begin{align}
C_1=\frac{e^{\frac{i\pi}{4}}}{2}(C_3+\frac{2C_4}{i}),\label{2.25a}\\
C_2=\frac{e^{-\frac{i\pi}{4}}}{2}(C_3-\frac{2C_4}{i}).\label{2.25b}
\end{align}
\end{subequations}

Also by defining $F_2=\frac{dV}{dr^*}|_{r=r_2}$ and changing the variable $u^{'}=(r^*-r_2){F_2}^{\frac{1}{3}}$, as in the previous case, it is obtained for $r\approx r_2$
\begin{multline}
\frac{|F_2|^{\frac{1}{6}}}{2\sqrt{\pi} \sqrt{|B_{sl}(r)|}}\Big{[}a^{'} e^{-\int_{r}^{r_2}|B_{sl}(r^*)| d r^*}+b^{'} e^{\int_{r}^{r_2}|B_{sl}(r^*)| d r^*}\Big{]}\\=\frac{C_3}{\sqrt{|B_{sl}(r)|}}\ e^{\int_{r_1}^{r} |B_{sl}(r^*)| d r^*}+\frac{C_4}{\sqrt{|B_{sl}(r)|}}\ e^{-\int_{r_1}^{r} |B_{sl}(r^*)| d r^*},\label{2.26}
\end{multline}
\begin{multline}
\frac{|F_2|^{\frac{1}{6}}}{\sqrt{\pi} \sqrt{B_{sl}(r)}}\Big{[}a^{'} \sin\big(\int_{r}^{r_1} B_{sl}(r^*) d r^*+\frac{\pi}{4}\big)+b^{'} \cos\big(\int_{r}^{r_1} B_{sl}(r^*) d r^*+\frac{\pi}{4}\big)\Big]\\=\frac{C_5}{\sqrt{B_{sl}(r)}}\ e^{i\int_{r_2}^{r} B_{sl}(r^*) d r^*}.\label{2.27}
\end{multline}
Now by using the following equations
\begin{subequations}
\begin{align}
&\gamma_{sl}(\omega)=\int_{r_1}^{r_2}|B_{sl}(r^*)| d r^*,\\
\int_{r_1}^{r} |B_{sl}&(r^*)| d r^*=\gamma_{sl}-\int_{r}^{r_2} |B_{sl}(r^*)|  d r^*,
\end{align}\label{2.28}
\end{subequations}
we define new coefficients
\begin{equation}
\begin{array}{lll}&  C_3^{'}=e^{\gamma_{sl}}\  C_{3},& \  C_4^{'}=e^{-\gamma_{sl}}\ C_4
\end{array}.\label{2.29}
\end{equation}
Equation \eqref{2.26} can be rewritten in terms of these new coefficients
\begin{multline}
\frac{|F_2|^{\frac{1}{6}}}{2\sqrt{\pi} \sqrt{|B_{sl}(r)|}}\Big[a^{'} e^{-\int_{r}^{r_2} |B_{sl}(r^*)| d r^*}+b^{'}\ e^{\int_{r}^{r_2} |B_{sl}(r^*)| d r^*}\Big]\\=\frac{C_3^{'}}{\sqrt{|B_{sl}(r)|}}\ e^{-\int_{r}^{r_2} |B_{sl}(r^*)| \ d r^*}+\frac{C_4^{'}}{\sqrt{|B_{sl}(r)|}}\ e^{\int_{r}^{r_2} |B_{sl}(r^*)|  d r^*},\label{2.30}
\end{multline}
and thus
\begin{equation}
C_3^{'}=\frac{a^{'}|F_2|^{\frac{1}{6}}}{2\sqrt{\pi}},\ \  C_4^{'}=\frac{b^{'}|F_2|^{\frac{1}{6}}}{\sqrt{\pi}}.\label{2.31}
\end{equation}
In a similar way for \eqref{2.27}, we get
\begin{equation}
C_5=\frac{|F_2|^{\frac{1}{6}}}{\sqrt{\pi}}\ e^{\frac{i\pi}{4}}b^{'},\ \  a^{'}=ib^{'}.\label{2.32}
\end{equation}
From \eqref{2.31} and \eqref{2.32}, the relationship between $C_4$ and $C_5$ can be found as
\begin{equation}
C_5=e^{\frac{i\pi}{4}}\ C_4^{'}=e^{\frac{i\pi}{4}}\ e^{-\gamma_{sl}}\ C_4.\label{2.33}
\end{equation}
Using \eqref{2.25a}, \eqref{2.25b}, and \eqref{2.33}, relation \eqref{2.34} is derived.

\end{document}